\documentclass[aps,prb,twocolumn,superscriptaddress,amsmath,amssymb]{revtex4-2}

\usepackage{amssymb}
\usepackage{graphicx}
\usepackage{tabularx}
\usepackage{bm}
\usepackage{epsfig}
\usepackage{graphicx}
\usepackage[ansinew]{inputenc}

\begin{document}

\title{The 3\textit{d} and 5\textit{d} electronic structures and orbital hybridization in Ba- and Ca-doped La$_2$CoIrO$_6$ double perovskite}

\author{J. R. L. Mardegan}
\affiliation{Deutsches Elektronen-Synchrotron DESY, Notkestra$\beta$e 85, 22607, Hamburg, Germany}

\author{L. S. I. Veiga}
\affiliation{Diamond Light Source, Chilton, Didcot, Oxfordshire, OX11 0DE, United Kingdom}

\author{T. Pohlmann}
\affiliation{Deutsches Elektronen-Synchrotron DESY, Notkestra$\beta$e 85, 22607, Hamburg, Germany}

\author{S. S. Dhesi}
\affiliation{Diamond Light Source, Chilton, Didcot, Oxfordshire, OX11 0DE, United Kingdom}

\author{S. Francoual}
\affiliation{Deutsches Elektronen-Synchrotron DESY, Notkestra$\beta$e 85, 22607, Hamburg, Germany}

\author{J. R. Jesus}
\affiliation{Centro Brasileiro de Pesquisas F\'{\i}sicas, 22290-180, Rio de Janeiro, RJ, Brazil}

\author{C. Macchiutti}
\affiliation{Centro Brasileiro de Pesquisas F\'{\i}sicas, 22290-180, Rio de Janeiro, RJ, Brazil}

\author{E. M. Bittar}
\affiliation{Centro Brasileiro de Pesquisas F\'{\i}sicas, 22290-180, Rio de Janeiro, RJ, Brazil}

\author{L. Bufai\c{c}al}
\email{lbufaical@ufg.br}
\affiliation{Instituto de F\'{\i}sica, Universidade Federal de Goi\'{a}s, 74001-970, Goi\^{a}nia, GO, Brazil}

\date{\today}

\begin{abstract}

Here we present a detailed investigation of the Co and Ir local electronic structures in La$_{1.5}A_{0.5}$CoIrO$_6$ ($A$ = Ba, Ca) compounds in order to unravel the orbital hybridization mechanism in these CoIr-based double perovskites. Our results of x-ray powder diffraction, ac and dc magnetization, Co and Ir $L_{2,3}$-edges and Co $K$-edge x-ray absorption spectroscopy and x-ray magnetic circular dichroism suggest a competition between magnetic interactions. A dominant antiferromagnetic coupling is found to be responsible for the ferrimagnetic behavior observed for $A$ = Ca below $\sim$96 K, the competing magnetic phases and the cationic disorder in this compound giving rise to a spin-glass state at low temperatures. For the $A$ = Ba, on the other hand, there is no evidence of long range order down to its spin-glass transition temperature. The remarkably different magnetic properties observed between these two compounds is discussed in terms of the structural distortion that alters the strength of the Co -- Ir couplings, with a relevant role played by the Co 3$d$ $e_g$ -- Ir 5$d$ $j_{eff}$ = 1/2 hybridization. 

\end{abstract}


\maketitle

\section{Introduction}

Since the discovery of ferrimagnetism (FIM) and half-metallic behavior above room temperature ($T$) in Sr$_2$FeMoO$_6$ \cite{Tokura}, double perovskites (DP) with various compositions have been produced and characterized in view of tuning the electronic and magnetic properties for spintronic applications. The most interesting properties are observed when the systems contain a combination of 3$d$ and 4$d$/5$d$ transition-metal (TM) ions \cite{Serrate,Sami}. Ir-based DPs have recently received considerable attention due to the peculiar interplay between the strong spin-orbit coupling (SOC), the Coulomb repulsion, and the crystal field splitting which, in octahedral coordination, lifts the Ir $t_{2g}$ orbital degeneracy into a fourfold $j_{eff}$ = 3/2 and a $j_{eff}$ = 1/2 doublet, giving rise to unusual magnetic properties \cite{Kim,Cao}.

The use of Co as 3$d$ TM in addition to 5$d$ TM adds further complexity to the stabilization of magnetism in these Ir-based DPs. A delicate balance between crystal field and interatomic exchange interactions can lead to different valences and spin states of the Co ions in the octahedral symmetry \cite{Raveau}. In particular, La$_2$CoIrO$_6$ shows exciting properties such as magnetodieletric effect \cite{Song} and cooperative octahedral breathing distortion \cite{Lee}. There is, however, some controversy concerning the microscopic mechanisms governing its magnetic properties.

N. Narayanan \textit{et al.}, from the analysis of neutron powder diffraction (NPD) data, ascribed the small ferromagnetic (FM)-like behavior in La$_2$CoIrO$_6$ to spin canting in the two interpenetrating antiferromagnetic (AFM) Co and Ir sublattices \cite{Narayanan}. However, the strong neutron absorption of Ir ions prevents a reliable determination of its magnetic structure. On the other hand, recent studies using x-ray absorption spectroscopy (XAS), x-ray magnetic circular dichroism (XMCD), and muon spin spectroscopy suggest an AFM coupling between Co and Ir FM sublattices, resulting in a ferrimagnetic (FIM) behavior \cite{Cho,PRB2020}. The small magnetization observed in the system can be ascribed to the antisite disorder (ASD) at Co/Ir sites and the competing magnetic interactions that lead to frustration. There are also other interpretations. Min-Cheol Lee \textit{et al.}, using Co and O $K$-edge XAS, claim that the spin frustration and extended paramagnetic (PM) phase are due to the non spin-selective nature of the Co$^{2+}$ (3$d$ $e_g\oplus4p$) -- Ir$^{4+}$ 5$d$ $e_g$ coupling \cite{Noh}. At the same time, recent Ir $L_3$-edge resonant inelastic x-ray scattering show the importance of the 3$d$-5$d$ hybridization in the magnetism of Ir-based DPs \cite{Jin}. Additionally, x-ray photoelectron spectroscopy measurements point to a large covalence in several DP iridates, questionning the description of these materials by a simple ionic picture \cite{Takegami}.

The observations reported above indicate how far we are from reaching a consensus on the nature of the magnetic ordering in Ir-based DPs. This is also valid when La is partially replaced by alkaline-earth in La$_2$CoIrO$_6$. For La$_{2-x}$Sr$_x$CoIrO$_6$, a structural transition accompanied by changes in the Co/Ir valences and magnetic structures is observed even for small Sr-doping \cite{Narayanan,Kolchinskaya}. For La$_{2-x}$Ca$_x$CoIrO$_6$, on the other hand, there is no structural transition, and up to 25\% of Ca to La, the substitution acts mainly at changing the Co valence, with Ir remaining nearly tetravalent \cite{PRB2020}. The $x$ = 0.5 concentration is of particular interest due to its compensation temperatures and spontaneous exchange bias effect \cite{Coutrim}, which seems to be directly related to a re-entrant spin glass (RSG) phase \cite{Model,Model2}. Likewise, the Sr to La partial substitution also induces competition between magnetic phases that leads to RSG \cite{SrCoIr}, whereas the La$_2$CoIrO$_6$ parent compound does not show glassy magnetic behavior.

In order to shed light on the role of the Co 3$d$ -- Ir 5$d$ orbital hybridization in stabilizing the magnetic coupling in CoIr-based DPs, we have performed a detailed investigation of how the crystal structure affects the electronic and magnetic structures of Co and Ir sublattices in La$_{1.5}$(Ba,Ca)$_{0.5}$CoIrO$_6$ compounds using x-ray powder diffraction (XRD), ac and dc magnetization, XAS and XMCD. Our results show remarkably different magnetic properties for these compounds, that are discussed in terms of the lattice distortions which ones alter the orbital hybridization between Co and Ir.

\section{Experimental details}

The polycrystalline samples here investigated were synthesized by conventional solid-state reaction. To obtain La$_{1.5}$Ba$_{0.5}$CoIrO$_6$ (hereafter called Ba0.5), stoichiometric amounts of La$_{2}$O$_{3}$, BaCO$_3$, Co$_{3}$O$_{4}$ and metallic Ir in powder form were mixed and heated at $650^{\circ}$C for 24 hours in air atmosphere. Later, the sample was re-grinded before a second step at $800^{\circ}$C for 48 hours. Finally, the material was grinded, pressed into a pellet, and heated at $1200^{\circ}$C for additional 48 hours. The synthesis route used to produce La$_{1.5}$Ca$_{0.5}$CoIrO$_6$ (herein called Ca0.5) is described elsewhere \cite{PRB2020}. 

High-resolution XRD data were collected at room temperature using a Cu $K_{\alpha}$ radiation operating at 40 kV and 40 mA at Centro Brasileiro de Pesquisas F\'{\i}sicas (CBPF), Brazil. The XRD data was carried over the angular range of $10^{\circ}\leq2\theta\leq110^{\circ}$, with a 2${\theta}$ step size of 0.01$^{\circ}$. Rietveld refinement was performed using GSAS software and its graphical interface program \cite{GSAS}. The ac and dc magnetic measurements were carried out using a Quantum Design PPMS magnetometer, at CBPF.

Low-temperature XAS measurements were carried out at the Co $K$- and Ir $L_{2,3}$-absorption edges on beamline P09 at PETRA III at the DESY \cite{Strempfer_JSR_2013} and at Co $L_{2,3}$-absorption edges on beamline I06 at Diamond Light Source (DLS). For the hard x-ray measurements, fine powder samples were mixed and pressed into a low-Z BN material to produce pellets for transmission measurements.  
XMCD measurements were performed at 5 K by fast-switching the beam helicity between left and right circular polarization \cite{Strempfer_AIP_2016}.  In order to align the magnetic domains and correct for nonmagnetic artifacts, an external magnetic field of $\pm$ 5 T was applied parallel and antiparallel to the incident beam wave vector $k$ using a 6T/2T/2T vector magnet. For the soft x-ray measurements, powder samples from the same batches were spaded on conductive carbon tapes and measured in total electron yield (TEY). The samples were cooled at 2 K, and an external magnetic field of $\pm$ 5 T was applied along the beam axis to reach saturation. 

\section{Results and discussion}

\subsection{x-ray diffraction}

\begin{figure}
\begin{center}
\includegraphics[width=0.45 \textwidth]{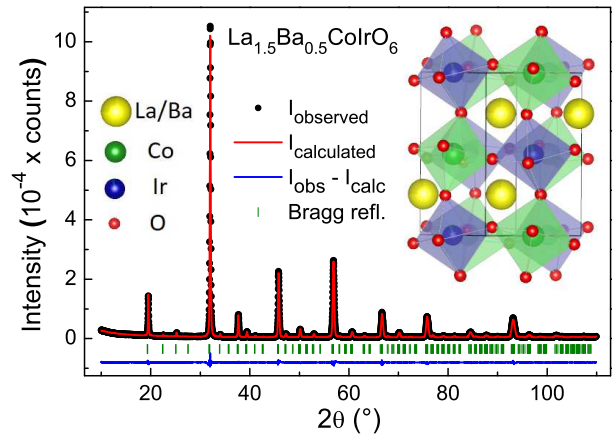}
\end{center}
\caption{Rietveld refinement fitting of Ba0.5 XRD. The vertical lines represent the Bragg reflections for the $P2_{1}/n$ space group. Inset shows the crystal structure, in which IrO$_6$ and CoO$_6$ are drawn as blue and green octahedra, respectively.}
\label{Fig_XRD}
\end{figure}

Fig. \ref{Fig_XRD} shows the XRD pattern of Ba0.5, which confirms the formation of a perovskite structure with no trace of impurity phase. Previous studies on La$_2$CoIrO$_6$ and LaBaCoIrO$_6$ show that the former compound grows in the monoclinic $P2_{1}/n$ space group whereas the latter was reported as belonging to the triclinic $I\overline{1}$ non-standard space group \cite{Narayanan,PRB2020,Battle}. Such non-standard setting is commonly adopted to facilitate the comparison of this structure with that of other DPs \cite{Sami,Woodward}. In our case, a detailed investigation of the XRD data of the Ba0.5 sample revealed that some of the experimentally observed reflections are not predicted for the triclinic symmetry. In contrast, all peaks match with that of the monoclinic structure. Therefore, the 25\% of Ba substitution at the La site is not sufficient to induce a structural transition, with Ba0.5 belonging to the $P2_{1}/n$ space group, the same as that of Ca0.5 \cite{PRB2020}. The main results obtained from the Rietveld refinement are displayed in Table \ref{T1}, where the structural parameters of Ca0.5 (some of them already reported in Ref. \onlinecite{PRB2020}) are also shown for comparison. 

From Table \ref{T1} we observe an increase in the unit cell volume ($V$) of Ba0.5 as compared to Ca0.5, in agreement with the Ba$^{2+}$ ionic radius being larger than Ca$^{2+}$ (respectively 1.61 and 1.34 $\AA$ in XII coordination \cite{Shannon}). Interestingly, although both compounds belong to the $P2_{1}/n$ monoclinic space group, there is a tendency towards a more tetragonal symmetry for Ba0.5, which is manifested by the closeness of the $a$ and $b$ lattice parameters and a significant increase in the $c$ parameter. The increased crystallographic symmetry is expected for larger rare-earth/alkaline-earth ionic radii in DPs \cite{Serrate,Sami}, manifested in the increased Co--O--Ir bond angles and the more similar Co--O and Ir--O bond lengths for Ba0.5. It must also be noticed in Table  \ref{T1} that the Co/Ir ASD, \textit{i.e.} the amount of Co(Ir) ions lying at Ir(Co) site, is rather similar for both compounds. Though this may look unexpected at a first glance, since an increased crystallographic symmetry usually leads to larger ASD, we recall that Ca0.5 was synthesized at $1000^{\circ}$C \cite{PRB2020} while Ba0.5 was produced at $1200^{\circ}$C. The synthesis condition was already shown to play a role on the B/B' cationic ordering of DPs, which commonly increases with increasing the synthesis temperature \cite{Hoffmann}.

\begin{table*}
\tiny
\caption{Main results obtained from the Rietveld refinements of the room temperature XRD data. Some of the parameters displayed for Ca0.5 were already reported in Ref. \onlinecite{PRB2020}.}
\label{T1}
\begin{tabular}{lllllllllllll}
\hline \hline
 & & & & & \multicolumn{5}{l}{Lattice Parameters and R factors} & & & \\
\hline
\multicolumn{2}{l}{Sample} & \multicolumn{2}{l}{$a$ (\AA)} & \multicolumn{2}{l}{$b$ (\AA)} & \multicolumn{2}{l}{$c$ (\AA)} & \multicolumn{2}{l}{$\beta$ ($^{\circ}$)} & $V$ (\AA$^{3}$) & $R_p$ (\%) & $R_{wp}$ (\%) \\
\hline

\multicolumn{2}{l}{Ca0.5} & \multicolumn{2}{l}{5.5566(1)} & \multicolumn{2}{l}{5.6427(2)} & \multicolumn{2}{l}{7.8808(2)} & \multicolumn{2}{l}{89.98(1)}  & 247.09(2) & 8.5 & 11.2 \\

\multicolumn{2}{l}{Ba0.5} & \multicolumn{2}{l}{5.6326(2)} & \multicolumn{2}{l}{5.6152(1)} & \multicolumn{2}{l}{7.9402(3)} & \multicolumn{2}{l}{90.02(1)} & 251.13(1) & 4.0 & 5.6 \\

\hline \hline
 & & & & & & \multicolumn{4}{l}{Bonds and Angles} & & & \\
\hline

Sample & Co-O$_1$ (\AA) & Co-O$_2$ (\AA) & Co-O$_3$ (\AA) & $\langle$Co-O$\rangle$ (\AA) & Ir-O$_1$ (\AA) & Ir-O$_2$ (\AA) & Ir-O$_3$ (\AA) & $\langle$Ir-O$\rangle$ (\AA) & Co-O$_1$-Ir ($^{\circ}$) & Co-O$_2$-Ir ($^{\circ}$) & Co-O$_3$-Ir ($^{\circ}$) & $\langle$Co-O-Ir$\rangle$ ($^{\circ}$) \\

\hline

Ca0.5 & 2.062(13) & 2.127(13) & 2.072(15) & 2.087(8) & 1.980(13) & 2.033(13) & 2.076(15) & 2.030(8) & 156.7(8) & 144.4(7) & 143.6(5) & 148.2(4) \\

Ba0.5 & 2.023(1) & 2.060(1) & 1.996(1) & 2.023(1) & 2.012(1) & 2.043(1) & 2.071(1) & 2.042(1) & 162.2(1) & 151.4(1) & 154.8(1) & 156.1(1) \\

\hline \hline

 & & & &  \multicolumn{5}{l}{Atomic Coordinates and Thermal Displacement Parameters} & & & \\
\hline

Sample & atom & \multicolumn{2}{l}{Wyckoff position} & \multicolumn{2}{l}{occupancy} & \multicolumn{2}{l}{x} & \multicolumn{2}{l}{y} & \multicolumn{2}{l}{z} & U$_{iso}$ (\AA$^{2}$) \\

\hline

 & La/Ca & \multicolumn{2}{l}{4$e$} & \multicolumn{2}{l}{0.75/0.25} & \multicolumn{2}{l}{0.5103(3)} & \multicolumn{2}{l}{0.5437(5)} & \multicolumn{2}{l}{0.2503(3)} & 0.0105(5) \\
 
 & Co1/Ir1 & \multicolumn{2}{l}{2$c$} & \multicolumn{2}{l}{0.927/0.073(3)} & \multicolumn{2}{l}{0} & \multicolumn{2}{l}{0.5} & \multicolumn{2}{l}{0} & 0.0121(3) \\

Ca0.5 & Ir2/Co2 & \multicolumn{2}{l}{2$d$} & \multicolumn{2}{l}{0.927/0.073(3)} & \multicolumn{2}{l}{0.5} & \multicolumn{2}{l}{0} & \multicolumn{2}{l}{0} & 0.0121(3) \\
 
  & O1 & \multicolumn{2}{l}{4$e$} & \multicolumn{2}{l}{1} & \multicolumn{2}{l}{0.2188(21)} & \multicolumn{2}{l}{0.2093(23)} & \multicolumn{2}{l}{0.9629(27)} & 0.0045(24) \\
  
    & O2 & \multicolumn{2}{l}{4$e$} & \multicolumn{2}{l}{1} & \multicolumn{2}{l}{0.3064(21)} & \multicolumn{2}{l}{0.7076(24)} & \multicolumn{2}{l}{0.9363(23)} & 0.0045(24) \\
    
      & O3 & \multicolumn{2}{l}{4$e$} & \multicolumn{2}{l}{1} & \multicolumn{2}{l}{0.3842(22)} & \multicolumn{2}{l}{0.9867(18)} & \multicolumn{2}{l}{0.2503(20)} & 0.0045(24) \\

\hline

 & La/Ba & \multicolumn{2}{l}{4$e$} & \multicolumn{2}{l}{0.75/0.25} & \multicolumn{2}{l}{0.5140(2)} & \multicolumn{2}{l}{0.5271(1)} & \multicolumn{2}{l}{0.2474(5)} & 0.0104(2) \\
 
 & Co1/Ir1 & \multicolumn{2}{l}{2$c$} & \multicolumn{2}{l}{0.936/0.064(2)} & \multicolumn{2}{l}{0} & \multicolumn{2}{l}{0.5} & \multicolumn{2}{l}{0} & 0.0129(2) \\

Ba0.5 & Ir2/Co2 & \multicolumn{2}{l}{2$d$} & \multicolumn{2}{l}{0.936/0.064(2)} & \multicolumn{2}{l}{0.5} & \multicolumn{2}{l}{0} & \multicolumn{2}{l}{0} & 0.0129(2) \\
 
 & O1 & \multicolumn{2}{l}{4$e$} & \multicolumn{2}{l}{1} & \multicolumn{2}{l}{0.2475(26)} & \multicolumn{2}{l}{0.2474(25)} & \multicolumn{2}{l}{0.9608(29)} & 0.0072(12) \\
  
 & O2 & \multicolumn{2}{l}{4$e$} & \multicolumn{2}{l}{1} & \multicolumn{2}{l}{0.3039(30)} & \multicolumn{2}{l}{0.6980(31)} & \multicolumn{2}{l}{0.9645(23)} & 0.0072(12)  \\
    
 & O3 & \multicolumn{2}{l}{4$e$} & \multicolumn{2}{l}{1} & \multicolumn{2}{l}{0.4214(28)} & \multicolumn{2}{l}{0.9996(11)} & \multicolumn{2}{l}{0.2548(29)} & 0.0072(12) \\

\hline \hline

\end{tabular}
\end{table*}

\subsection{ac and dc magnetization}

\begin{figure*}
\begin{center}
\includegraphics[width= \textwidth]{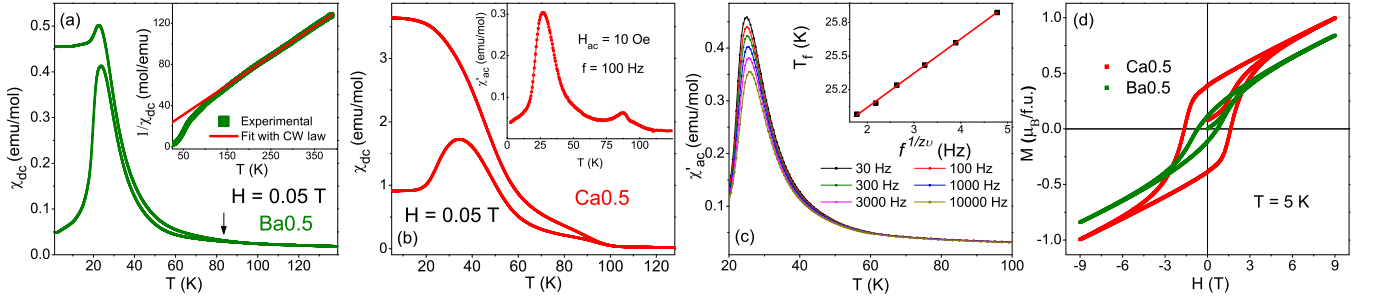}
\end{center}
\caption{(a) ZFC-FC $\chi_{dc}$ curves for Ba0.5, measured with $H$ = 0.05 T. The arrow indicates the temperature where the curves bifurcate, and the inset shows the best fit of PM region with the CW law. (b) ZFC-FC $\chi_{dc}$ curves measured with $H$ = 0.05 T for Ca0.5, adapted from Ref. \onlinecite{PRB2020}. Inset shows the $\chi'_{ac}$ curve measured with $H_{ac}$ = 10 Oe, $f$ = 100 Hz. (c) $\chi'_{ac}$ as a function of $T$ for Ba0.5 at different $f$, measured with $H_{ac}$ = 10 Oe. Inset shows $T_f$ as a function of $f$, where the solid line represents the best fit to the data using Eq. \ref{EqSG1}. (d) M(H) curves for Ba0.5 and Ca0.5 samples, measured at 5 K.}
\label{Fig_MxT}
\end{figure*}

Fig. \ref{Fig_MxT}(a) shows the  zero-field cooled (ZFC) and field cooled (FC) dc magnetic susceptibility ($\chi_{dc}$) data for Ba0.5, measured with a magnetic field ($H$) of 0.05 T. For comparison, Fig. \ref{Fig_MxT}(b) displays $\chi_{dc}$ curves for Ca0.5 also measured with $H$ = 0.05 T, as reported in Ref. \onlinecite{PRB2020}. For Ca0.5, the higher-$T$ FM-like magnetic transition at $T\sim96$ K is ascribed to Co$^{2+}$--O--Ir$^{4+}$ AFM coupling that results in FIM. At the same time, the peak at $T\sim27$ K is related to the emergence of a spin glass (SG) phase caused by competing magnetic interactions and ASD \cite{PRB2020}. 

For Ba0.5, although there is only one clear peak at $T\sim24$ K in the $\chi_{dc}$ data, a bifurcation of the ZFC and FC curves at $T\sim85$ K is observed, at which temperature a deviation of the Curie-Weiss law is also observed in the inverse susceptibility [inset of Fig. \ref{Fig_MxT}(a)]. This could indicate that a non-percolating short-range magnetic coupling starts to develop around this temperature, which could be in turn related to the appearance of a spin-glass phase at much lower temperatures. The emergence of short-range correlations and fluctuations far above the freezing temperature is commonly observed in spin-glasses \cite{Morgownik}.

The real part of the ac magnetic susceptibility ($\chi'_{ac}$) curves measured for Ba0.5 in an oscillating field $H_{ac}$ = 10 Oe with six ac frequencies ($f$) ranging from 30 to 10000 Hz is shown on Fig. \ref{Fig_MxT}(c). In contrast to the ac magnetic susceptibility data for Ca0.5, where a clear cusp is observed at $T_C$ [inset of Fig. \ref{Fig_MxT}(d)], the $\chi'_{ac}$ curves of Ba0.5 gives no evidence of any anomaly close to 85 K, further indicating that the weak magnetic coupling between Co and Ir does not percolate on this compound. At lower temperatures, it is observed a systematic decrease in the amplitude of the $T\sim24$ K peak and its shift to higher temperatures as $f$ increases. These are characteristic features of glassy magnetism, signaling that this peak corresponds to the freezing temperature ($T_f$) of a SG-like phase \cite{Mydosh}. The evolution of $T_f$ with $f$ turns out to be well fitted by the power-law equation of the dynamic scaling theory, commonly used to investigate SG-like systems
\begin{equation}
\frac{\tau}{\tau_{0}}=\left[\dfrac{(T_{f} - T_{g})}{T_{g}}\right]^{-z\nu}, \label{EqSG1}
\end{equation}
where $\tau$ is the relaxation time corresponding to the measured frequency, $\tau_{0}$ is the characteristic relaxation time of spin flip, $T_{g}$ is the SG transition temperature ($T_f$ as $f$ tends to zero), $z$ is the dynamical critical exponent and $\nu$ is the critical exponent of the correlation length \cite{Mydosh,Souletie}. The solid line on inset of Fig. \ref{Fig_MxT}(c) represents the fit to Eq. \ref{EqSG1}, yielding $T_{g}\simeq24.4$ K, $\tau_{0}\simeq6.3\times10^{-12}$ s and $z\nu\simeq5.9$, being these values typical of conventional SG systems \cite{Coutrim,Souletie,Mydosh2}.

Usually, the relative shift $\delta T_f$ = $\Delta T_f/T_f(\Delta ln f)$ is calculated to classify the material as canonical SG, cluster SG or superparamagnet. For Ba0.5 we find $\delta T_f\simeq$0.006, which is within the range usually found for canonical SG ($\delta T_f$ $<$ 0.01 \cite{Mydosh2,Malinowski,Anand2}). This confirms that, as for Ca0.5, the ASD and competing magnetic interactions also lead to the emergence of a SG phase in Ba0.5.

The fit of the PM region of 1/$\chi_{dc}$ raw data with the Curie-Weiss (CW) law, inset of Fig. \ref{Fig_MxT}(a), yields a CW temperature $\theta_{CW}=-58$ K for Ba0.5. The negative sign indicates that AFM couplings dominate the magnetic interactions in this compound, as observed for Ca0.5 \cite{PRB2020}. The $\mu_{eff}=5.3$ $\mu_B$/f.u. obtained from the fit is fairly close to the value found for Ca0.5 (5.4 $\mu_B$/f.u.) \cite{PRB2020}. It is important to mention that the situation here is quite different from that of some Ir$^{5+}$-based compounds where a $\chi_0$ term is usually included in the CW equation to account for a diamagnetic contribution to the susceptibility \cite{Cao,Dey,Agrestini,Wolff}. In our Ir$^{4+}$/Co$^{2+,3+}$ system, the van-Vleck constant term and the core diamagnetic contribution are expected to be negligible regarding the paramagnetic moments of the TM ions.

Fig. \ref{Fig_MxT}(d) displays the magnetization as a function of $H$ [M(H)] curves measured at 5 K for Ba0.5 and Ca0.5 samples. Both curves are typical of FIM systems, with a FM-like hysteresis but a linear increase of magnetization at high $H$, further indicating the presence of AFM interactions. It is interesting to notice the remarkable decrease in the coercive field ($H_C$) of Ba0.5 with respect to Ca0.5. This resembles other 3$d$-5$d$ DPs for which $H_C$ dramatically decreases with the expansion of the unit cell \cite{Sikora,Haskel,Ferreira,Feng}.

\subsection{XAS and XMCD measurements}

\begin{figure}
\begin{center}
\includegraphics[width=0.55 \textwidth]{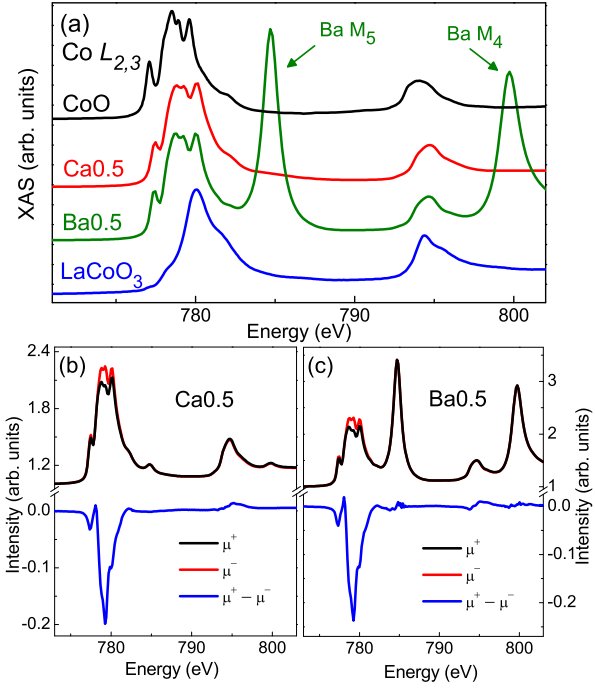}
\end{center}
\caption{(a) Co $L_{2,3}$-edge XAS spectra of Ca0.5 and Ba0.5 samples, together with those of CoO (reference for Co$^{2+}$) and LaCoO$_3$ (reference for Co$^{3+}$). The presence of Ba $M_{4,5}$-edges in Ba0.5 sample is highlighted by green arrows. (b) and (c) show respectively the XMCD spectra of Ca0.5 and Ba0.5  samples, taken at 2 K with $H$ = 5 T. The photon spin was aligned parallel ($\mu^{+}$, black) and antiparallel ($\mu^{-}$, red) to $H$, and the difference spectra are shown in blue.}
\label{Fig_CoL}
\end{figure}

XAS spectra at the TM $L_{2,3}$-edges are very sensitive to the valence state, corresponding to electronic transitions from the 2$p$ to 3$d$ (5$d$) orbitals for ions of the first (third) row.Fig. \ref{Fig_CoL}(a) shows the Co $L_{2,3}$-edge spectra for the samples of interest together with those of CoO and LaCoO$_3$, used here as standards for Co$^{2+}$ and Co$^{3+}$, respectively. The presence of Co$^{2+}$ in Ba0.5 and Ca0.5 samples is confirmed by the similarity of the spectra with that of CoO. However, for Ba0.5 it is observed the presence of pronounced peaks at $\sim$785 and $\sim$800 eV, associated with Ba $M_{4,5}$ absorption edges. A closer inspection of the higher energy side of the $L_3$-edge ($\sim$780 eV) of Ca0.5 and Ba0.5 reveals a relative increase in the intensity of the spectra with respect to that of CoO near the energy position of Co$^{3+}$ $L_3$-edge absorption, suggesting that some fraction of Co is in trivalent state. This is in agreement with the effective magnetic moments obtained for both compounds \cite{PRB2020}.

A quantitative estimate of the Co oxidation state can be made for our Ca0.5 sample by comparing the \textit{center of mass} of its $L_{3}$-edge spectrum with those of the reference samples. Assuming a linear variation in energy between the \textit{center of mass} of Co$^{2+}$ and Co$^{3+}$, the Co formal valence was estimated to be $\sim$2.4 for Ca0.5, which would mean the presence of $\sim$10\% of Ir$^{5+}$ or oxygen vacancy in this sample. Surely this is just a rough estimate, since other parameters such as the crystal field and structure may play a role on the integrated intensity of XAS curves. Nevertheless, a +2.4 formal valence is reasonably close to the +2.5 value expected for a 25\%  Ca$^{2+}$ to La$^{3+}$ substitution, which one would act mainly at changing the Co valence, a scenario corroborated by the magnetometry results and Ir $L_{3}$-edge XAS. For Ba0.5, the presence of the Ba-$M_{5}$ absorption peak nearby the Co $L_{3}$-edge curve affects the calculation of the \textit{center of mass}, preventing a reliable estimate of its valence state. Notwithstanding, the similarity of its $L_{2,3}$-edge curves with that of Ca0.5, as well as the closeness of the $\mu_{eff}$ observed for both samples, suggest that the Co valence for Ba0.5 is similar to that of Ca0.5.

Figs. \ref{Fig_CoL}(b) and (c) show the Co $L_{2,3}$ XMCD of Ca0.5 and Ba0.5 samples, respectively. The same overall behavior suggests nearly the same magnetic structure for both compounds. The black and red curves stand respectively for $\mu^{+}$ and $\mu^{-}$, \textit{i.e.}, for parallel and antiparallel alignments between the photon spin and $H$, and the difference spectra $\Delta\mu = \mu^{+} - \mu^{-}$ corresponds to the blue line. As can be noticed, the XMCD signal is largely negative at $L_3$-edge but with a very small positive value at $L_2$-edge, indicating non-negligible orbital contribution to the magnetic moment, as commonly observed for Co in octahedral coordination \cite{Raveau}. For a quantitative analysis of the Co-$L_{2,3}$ XMCD spectra, we have used the sum rules to derive the orbital and spin contributions to the magnetization \cite{Thole,Carra}
\begin{equation}
m_{l}=- \frac{4\int_{L_3+L_2}(\mu^+-\mu^-)d\omega}{3\int_{L_3+L_2}(\mu^++\mu^-)d\omega}N_{h}, \label{Eq9}
\end{equation} 
\begin{equation}
\begin{split}
m_{s} & = -\frac{6\int_{L_3}(\mu^+-\mu^-)d\omega-4\int_{L_3+L_2}(\mu^+-\mu^-)d\omega}{\int_{L_3+L_2}(\mu_++\mu_-)d\omega}\times \\
& N_{h}\left(1+ \frac{7\langle T_z\rangle}{2\langle S_z \rangle}\right)^{-1},  \label{Eq10}\\
\end{split}
\end{equation} 
where $m_{l}$ and $m_{s}$ are the angular and spin magnetic moments in units of $\mu_B$/atom, $S_z$ denotes the projection along $z$ of the spin magnetic momentum, $N_h$ represent the number of empty 3$d$ states, $T_z$ denotes the magnetic dipole moment, and $L_2$ and $L_3$ represent the integration range. Here we used $N_h$ = 3.5$\pm$0.5, which is an approximated atomistic value corresponding to a mixture of Co$^{2+}$ ($Nh$ = 3) and Co$^{3+}$ ($Nh$ = 4). Moreover, $T_z$ is estimated to be negligible against $S_z$ for TM ions in octahedral symmetry \cite{Teramura,Groot}. With these considerations, we obtained $m_{l}$ = 0.15 $\mu_B$ and $m_{s}$ = 0.29 $\mu_B$. Considering sources of deviations such as electronic interactions, imprecision in the integral calculations, and the $N_h$ value assumed, we estimate an uncertainty of $\sim20\%$ on these values \cite{Teramura,Groot}. 

The calculated moments are similar to those reported for La$_2$CoIrO$_6$ ($m_{l}$ = 0.18 $\mu_B$, $m_{s}$ = 0.31 $\mu_B$) \cite{Cho}, which is somewhat unexpected considering that from our XAS spectra we can expect about 40\% of the Co ions in a trivalent state, as opposed to a presumably 100\% of Co$^{2+}$ in La$_2$CoIrO$_6$. However, in Ref. \onlinecite{Cho} a quantitative investigation of the Co formal valence was not performed for the parent compound, whilst another study performed on a La$_2$CoIrO$_6$ polycrystalline sample produced by a similar synthesis route reported the presence of some amount of Co$^{3+}$ \cite{PRB2020}. Furthermore, the previous XMCD study reported for La$_2$CoIrO$_6$ in Ref. \onlinecite{Cho} was carried out at 43 K using $H$ = 0.8 T, while element-specific hysteresis curves measured for CoIr-DPs have already shown that Co is far from saturation at these conditions \cite{Kolchinskaya}. 

For the Ba0.5 sample, one must recall the pronounced Ba $M_{4,5}$-edge peaks observed near the Co $L_{2,3}$ absorption edge. Although one may not expect a XMCD signal at the Ba $M_{4,5}$-edge, the large XAS white line intensity may influence the background of the Co $L_{2,3}$ pre- and post-edge regions, and affect the adjustment of the XAS curve for the calculation of the $L_2 + L_3$ integral. Therefore, $m_{l}$ and $m_{s}$ were not estimated for this compound. The $m_{l}$/$m_{s}$ ratio, however, is independent of the XAS integrated intensity, resulting in a trustworthy value that is fairly close to that found for Ca0.5.

\begin{table*}
\caption{Co and Ir orbital ($m_l$), spin ($m_s$) and total ($m_t$) magnetic moments obtained from the sum rules calculations of Co and Ir $L_{3,2}$ XMCD spectra.}
\label{T2}
\begin{tabular}{c|cccc|cccc}
\hline \hline

Sample & Co $m_l$ ($\mu_B$) & Co $m_s$ ($\mu_B$) & Co $m_t$ ($\mu_B$) & Co  $m_l/m_s$ & Ir $m_l$ ($\mu_B$) & Ir $m_s$ ($\mu_B$) & Ir $m_t$ ($\mu_B$) & Ir  $m_l/m_s$\\

\hline

Ca0.5 &  0.15$\pm$0.03 & 0.29$\pm$0.06 & 0.44$\pm$0.07 & 0.52$\pm$0.11 & -0.11$\pm$0.02 & -0.11$\pm$0.02 & -0.22$\pm$0.03 & 1.00$\pm$0.21 \\

Ba0.5 &  - & - & - & 0.54$\pm$0.15 & -0.07$\pm$0.01 & -0.06$\pm$0.01 & -0.13$\pm$0.01 & 1.17$\pm$0.27 \\

\hline \hline

\end{tabular}
\end{table*}

\begin{figure}
\begin{center}
\includegraphics[width=0.43 \textwidth]{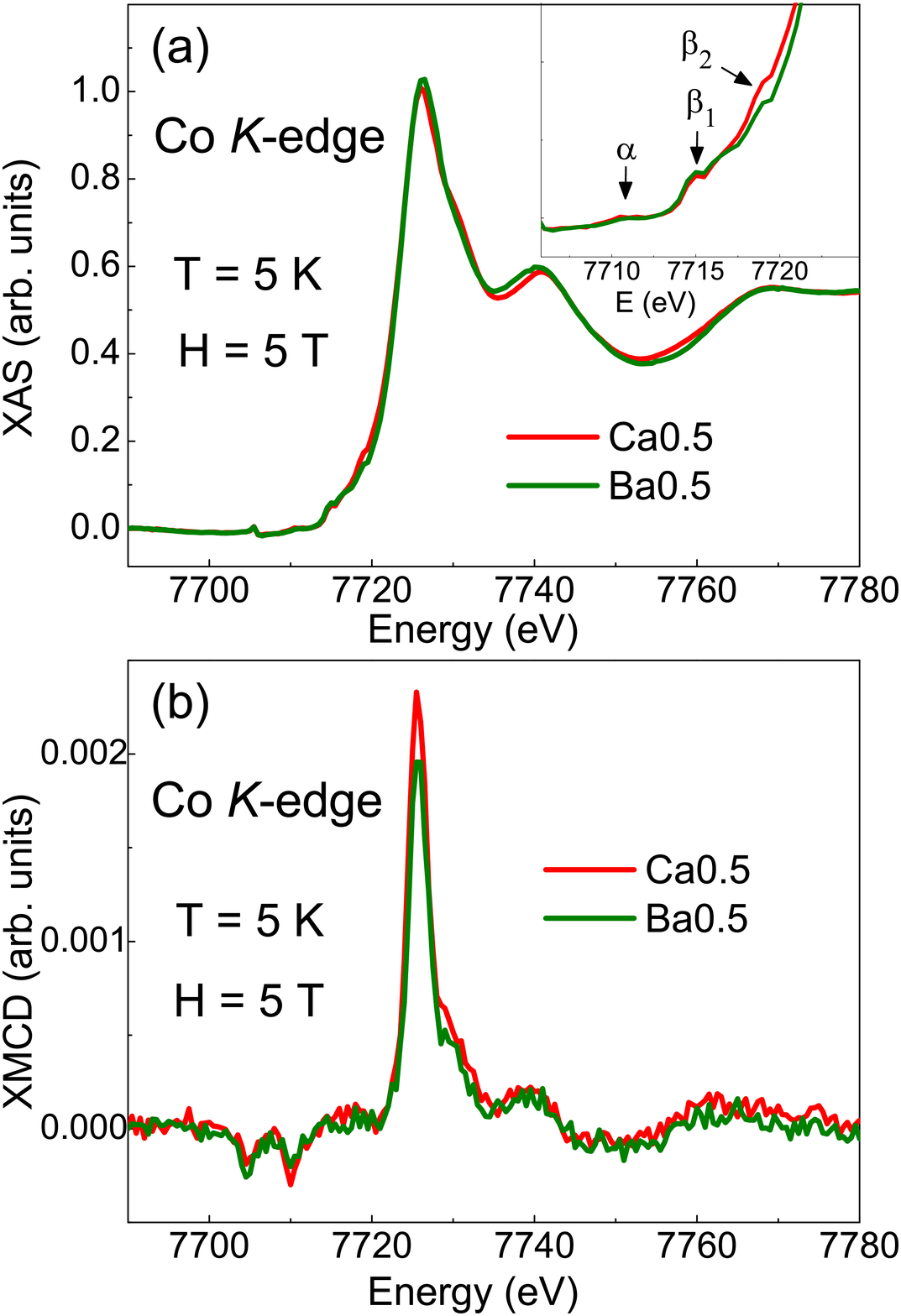}
\end{center}
\caption{(a) Co $K$-edge XAS spectra of Ba0.5 and Ca0.5 samples, measured at 5 K. Inset shows a magnified view of the pre-edge region, highlighting the $\alpha$, $\beta_1$ and $\beta_2$ peaks.  (b) The XMCD spectra of these samples.}
\label{Fig_CoK}
\end{figure}

In order to avoid the problem of Ba $M_{4,5}$-edges, we investigated the Co $K$-edge XAS and XMCD, as shown in Fig. \ref{Fig_CoK}. The Co $K$-edge XAS reflects mainly the electric dipole transitions from 1$s$ to 4$p$ level, where the edge position in energy may probe the Co valence. At the same time, local arrangements of neighboring ions determine the region after the absorption threshold. The similarity between the curves displayed in Fig. \ref{Fig_CoK}(a) is in accordance with the same monoclinic structure and similar Co valence states for both compounds.

\begin{figure}
\begin{center}
\includegraphics[width=0.6 \textwidth]{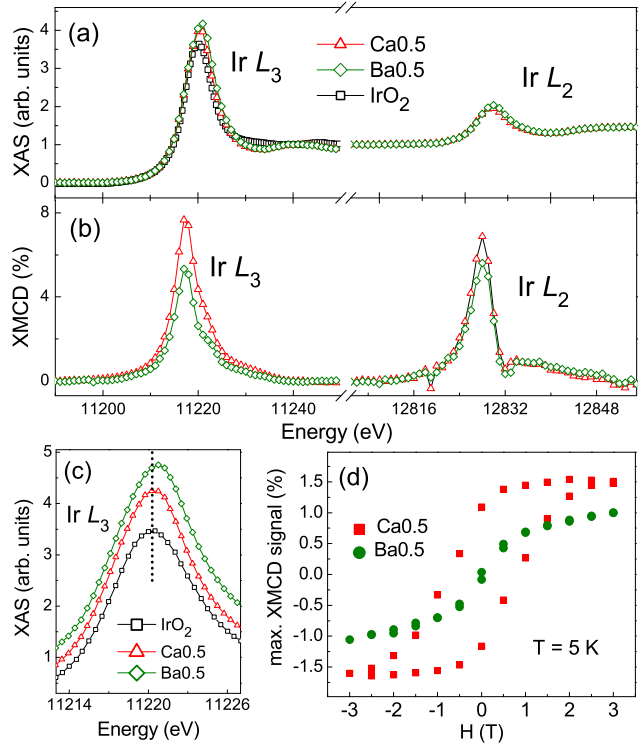}
\end{center}
\caption{Ir $L_{2,3}$-edges: (a) XAS and (b) XMCD spectra for Ba0.5 and Ca0.5 measured at 5 K and under 5 T. (c) Magnified view of the $L_{3}$-edge, where the IrO$_2$ standard is added for comparison. (e) Element-specific XMCD hysteresis loops of Ir $L_3$-edge.}
\label{Fig_IrL}
\end{figure}

The inset of Fig. \ref{Fig_CoK}(a) shows a magnified view of the pre-edge region. Although the $K$-edge XAS reflects mostly the transitions to unoccupied Co 4$p$ states, M.-C. Lee \textit{et al.} showed that the pre-edge anomalies $\alpha$ and $\beta_1$ are related to Co 4$p$ partial electron filling via intra-site mixing with Co 3$d$ orbitals \cite{Noh}. Thus, the orbital hybridization with Ir 5$d$ orbitals may largely contribute to these pre-edge features (see Fig. \ref{Fig_Draw}). Interestingly, a third anomaly, $\beta_2$ $\sim7720$ eV, is observed for Ca0.5 and Ba0.5, which may be related to Co$^{3+}$ (3$d$ $e_g\oplus4p$) -- Ir 5$d$ $e_g$ hybridization via O 2$p$.

The Co $K$-edge XMCD signals measured at $T$ = 5 K and $H$ = 5 T are very similar for both Ca0.5 and Ba0.5 [Fig. \ref{Fig_CoK}(b)], suggesting that although the Co 3$d$ bands present a relative extended character, the changes in the crystallographic environment does not drastically affect the Co magnetism.

Similar scenario cannot be extrapolated to the Ir ions, since the 5$d$ orbitals are even more extended than the Co 3$d$, which makes Ir ions more sensitive to structural and electronic changes. Fig. \ref{Fig_IrL}(a) shows the Ir $L_{2,3}$-edge XAS spectra of Ba0.5 and Ca0.5 samples together with that of IrO$_2$, reference for Ir$^{4+}$ state. The spectra of Ba0.5 and Ca0.5 are reasonably similar to that of IrO$_2$, indicating that the majority of Ir ions are in a tetravalent state for both compounds. Nonetheless, the similarity between the curves is characteristic of the $L_{2,3}$-edge spectra of 5$d$ TM ions, being related to the more diffuse valence orbitals compared to the 3$d$ ones \cite{Kolchinskaya,Liu}. A magnified view of the $L_3$-edge region depicted in Fig. \ref{Fig_IrL}(c) indicates some subtle differences between the spectra. The Ca0.5 and Ba0.5 spectra are shifted of respectively $\sim$0.2 eV and $\sim$0.3 eV with respect to the IrO$_2$ $L_3$ XAS, suggesting the presence of a small amount of Ir$^{5+}$. However, these shifts lie within the instrumental resolution ($\sim$1.5 eV), while the difference in energy between the white line positions of Ir$^{4+}$ and Ir$^{5+}$ is of about 1.2 eV \cite{Agrestini}. With these considerations, we can only conclude that Ba0.5 and Ca0.5 presents nearly the same Ir valence state, with the majority of ions in a tetravalent state.

Fig. \ref{Fig_IrL}(b) shows the Ir $L_{2,3}$-edge XMCD spectra measured at $T$ = 5 K and $H$ = 5 T. The positive signs of the Ir XMCD signals, combined with the negative sign for the Co signal, indicate that those two ions develop an AFM coupling in both compounds. Table \ref{T2} shows the Ir orbital and spin moments obtained from the sum rules, for which it was assumed $N_h$ = 5$\pm$0.5 (Ir$^{4+}$). The $m_l$ and $m_s$ values found are similar to those previously reported for La$_{1.5}$Sr$_{0.5}$CoIrO$_{6}$ \cite{Kolchinskaya}, and the large $m_l$/$m_s$ ratio observed is typical of iridates and other 5$d$-based materials presenting strong SOC \cite{Yi,Fujiyama}. The somewhat smaller moments observed for Ba0.5 with respect to Ca0.5 helps to explain its slightly smaller $\mu_{eff}$ obtained from the $\chi_{dc}$ curves, also suggesting that the Ir magnetization is closely related to its orbital hybridization with neighboring Co ions.

Fig. \ref{Fig_IrL}(c) shows element-specific hysteresis loops obtained by monitoring the Ir $L_3$-edge XMCD signal as a function of $H$. As can be noticed, the Ir moment is nearly saturated for Ca0.5 under $H$ = 3 T, whereas for Ba0.5 the saturation might be reach only for much higher magnetic fields. A remarkable point in these curves is the pronounced decrease in the $H_C$ of Ba0.5 with respect to Ca0.5, as also observed in the macroscopic M(H) curves measured in PPMS [Fig. \ref{Fig_MxT}(d)]. 

\subsection{Discussion}

\begin{figure}
\begin{center}
\includegraphics[width=0.6 \textwidth]{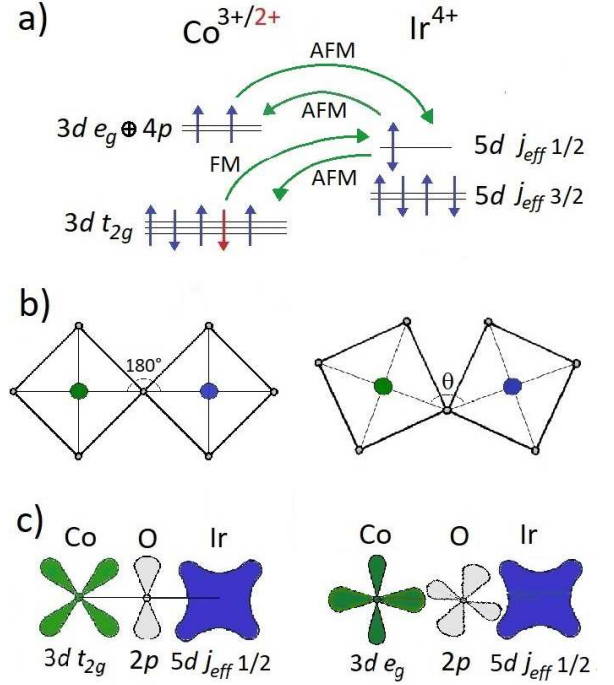}
\end{center}
\caption{(a) Schematic diagram of the mechanisms of hybridization between Co$^{2+/3+}$ (3$d^7$/3$d^6$) and Ir$^{4+}$ (5$d^5$), where the red arrow represents an additional spin down electron for Co$^{2+}$ $t_{2g}$, which is absent for Co$^{3+}$.  The  green arrows represent virtual electron hopping. The blue double-arrow in Ir $j_{eff}$ = 1/2 is spin down for the case of AFM interaction and spin up for the FM case. (b) (Co/Ir)O$_6$ octahedra geometries for Co--O--Ir bond angle $\theta$ = $180^{\circ}$ and $\theta$ $<$ $180^{\circ}$. (c) The Co $t_{2g}$ -- Ir $j_{eff} = 1/2$ (left) and Co (3$d$ $e_g\oplus4p$) -- Ir $j_{eff} = 1/2$ (right) orbital hybridizations.}
\label{Fig_Draw}
\end{figure}

In general, the magnetic ordering of TM ions in DPs can be described by the orbital hybridization mechanism, according to which the $d$ orbital states reside close to the Fermi surface, being respectively slightly below the chemical potential for one TM ion and slightly above for another. Then, a hybridization between these orbitals, mediated by O 2$p$ orbitals, enables the virtual spin-dependent electron hopping from one TM to another and can stabilize either FM or AFM coupling \cite{Serrate,Sami}. For HS Co$^{2+}$ (Co$^{3+}$), there are two (one) completely filled and one (two) half filled $t_{2g}$ orbital(s), while both $e_g$ orbitals are partially filled. The 5$d^5$ Ir$^{4+}$ ion, on the other hand, is at LS configuration, with the SOC lifting the $t_{2g}$ degeneracy to a fourfold $j_{eff}$ = 3/2 and a twofold $j_{eff}$ = 1/2 state. Thus, Ir$^{4+}$ has one $j_{eff}$ = 1/2 electron while its $j_{eff}$ = 3/2 orbitals are fully occupied \cite{Kim,Jackeli}. 

Recent first-principles calculations for La$_2$CoIrO$_6$ have shown that the energy of the Ir-$j_{eff}$ = 1/2 state resides in between those of Co $t_{2g}$ and $e_g$, whereas the strong crystal field splitting on Ir puts its $e_g$ states $\sim$3 eV higher in energy, making these later orbitals irrelevant to our discussion \cite{Ganguly}. As a consequence, only the $t_{2g}$ -- O 2$p$ -- $t_{2g}$ $\pi$ bonding would be a relevant exchange path between Co and Ir. However, although the $t_{2g}$ -- O 2$p$ -- $e_g$ coupling is in principle forbidden by symmetry in a cubic perovskite (\textit{i.e.} for a 180$^{\circ}$ B--O--B' bond angle), it can become possible with lattice distortion, as is the case for the samples here investigated. 

Fig. \ref{Fig_Draw}(a) depicts the spin-selective exchange pathways between Co$^{2+}$/Co$^{3+}$ and Ir$^{4+}$ in our system. There is one possible FM virtual hopping from the Co $t_{2g}$ minority spin state to Ir $t_{2g}$ ($j_{eff}$ = 1/2), and a number of possible AFM couplings. Here we must recall the highly covalent character of the Ir--O bond, with an O-electron partially occupying the Ir-$t_{2g}$ orbital \cite{Takegami}, which in principle inhibits the virtual electron transfer from Co to Ir. This partial population of Ir-$j_{eff}$ = 1/2 orbital with O-electrons, together with the SOC, may explain the dramatic decrease of the Ir magnetic moment observed here and for other Ir$^{4+}$-based perovskites \cite{Narayanan,Kolchinskaya,Lee,PRB2020}. Therefore, AFM charge transfer from Ir to Co seems to dominate the orbital hybridization in these samples, which is supported by the fact that all experimental and theoretical works reported so far for CoIr-based DPs points towards AFM coupling between these ions. As Fig. \ref{Fig_Draw}(a) shows, the intermediate charge transfer from Ir $t_{2g}$ can occur to a Co $t_{2g}$ or $e_{g}$ orbital. A transfer to Co $t_{2g}$, through the Ir $t_{2g}$ -- O 2$p$ -- Co $t_{2g}$ pathway, is expected to be stronger for Ba0.5 than for Ca0.5, since the former compound presents increased Co--O--Ir bond angles [see Figs. \ref{Fig_Draw}(b) and (c)]. The fact that Ca0.5 shows a clear magnetic ordering at a relatively high temperature, contrasting the results found for Ba0.5, suggests that the Ir $t_{2g}$ -- O 2$p$ -- Co $t_{2g}$ is not the most relevant hybridization channel in these compounds. On the other hand, the Ir $t_{2g}$ -- O 2$p$ -- Co $e_g$ channel  is expected to strengthen as the Co--O--Ir angle decreases, subsequently becoming the most relevant hybridization pathway between Co and Ir.  This is reminiscent of other 3$d$-5$d$ DPs, for which a systematic increase of the magnetic ordering temperature is observed as the B--O--B' angle decreases \cite{Sikora,Ferreira,Feng,Morrow}. The absence of a clear magnetic ordering for Ba0.5 is thus likely related to the weakened Ir $t_{2g}$ -- O 2$p$ -- Co $e_g$ and strengthened Ir $t_{2g}$ -- O 2$p$ -- Co $t_{2g}$ hybridization, with the competing magnetic interactions leading to spin frustration.

Another striking difference between Ba0.5 and Ca0.5 resides in their $H_C$. In other 3$d$-5$d$ DPs, as well, changes in the orbital hybridization caused by structural distortions remarkably alter their magnetic properties. For $AA'$FeReO$_6$ ($AA'$ = Ba, Ca, Sr), the increase of $H_C$ and $T_C$ was observed with decreasing the average $A$ radius, which was ascribed to the increase of Re orbital moment due to lattice distortion \cite{Sikora}. Contrastingly, high-pressure studies in $A_2$FeReO$_6$ ($A$ = Ba, Ca) showed a remarkable increase of $H_C$ with increasing pressure that was not related to the increase of Re orbital contribution. Instead, such changes in $H_C$ were explained in terms of reducing the strength of SOC caused by the structural distortion \cite{Haskel}. 

For our CoIr-based samples, the dramatic increase in the $H_C$ of Ca0.5 with respect to Ba0.5 is probably not related to increase of the orbital moments since the Ir $m_l$ value obtained for Ca0.5 is only slightly larger than that of Ba0.5, while the Co $m_l/m_s$ ratio is higher for the latter compound. Here we must recall that FeRe-based DPs are somewhat different from our CoIr-system. Re$^{5+}$ has two unpaired electrons at $j_{eff}$ = 3/2 orbitals whereas for Ir$^{4+}$ these orbitals are complete and there is one unpaired spin at $j_{eff}$ = 1/2. Moreover, a quenched orbital moment is expected for Fe$^{3+}$, whereas Co$^{2+}$ is known to present a non-negligible orbital contribution, as our XMCD results clearly indicate. Our results may find more resemblance with those observed for NiIr-based DPs, since Ni$^{2+}$ may also present orbital contribution to the magnetic moment \cite{Giles,Kwon}. For $A_2$NiIrO$_6$ ($A$ = La-Lu) the systematic increase of $T_C$ and $H_C$ with decreasing $A$ ionic radius is attributed to the increased strength of the AFM Ni$^{2+}$ $e_g$ -- Ir$^{4+}$ $t_{2g}$ orbital hybridization \cite{Ferreira,Feng}. The enhancement of $H_C$ with bending of B--O--B' angle in the La$_{2-x}$A$_x$CoIrO$_6$ samples here investigated gives further evidence that the $t_{2g}$ -- O 2$p$ -- $e_g$ hybridization plays a major role on the CoIr-based DPs.

\section{Summary}

In summary, we thoroughly investigated the structural, electronic and magnetic properties of La$_{1.5}$Ba$_{0.5}$CoIrO$_{6}$ (Ba0.5) and La$_{1.5}$Ca$_{0.5}$CoIrO$_{6}$ (Ca0.5) samples by means of XRD, ac and dc magnetization, Co and Ir $L_{2,3}$-edges and Co $K$-edge XAS and XMCD. Our results show that the presence of 25\% of Ba$^{2+}$/Ca$^{2+}$ at La$^{3+}$ site leads to Co$^{2+}$/Co$^{3+}$ mixed valence, while Ir remains mostly in a tetravalent state. For Ca0.5 it is observed a magnetic ordering at $\sim$96 K and a large $H_C$, while for Ba0.5 a magnetic ordering is not clear and the $H_C$ is much smaller. Since both compounds presents nearly the same Co and Ir valence states and effective magnetic moments, the remarkable differences between the magnetic properties of these samples can be understood in terms of changes in the exchange paths between Co $e_g$/$t_{2g}$ and Ir $t_{2g}$ ($j_{eff}$ = 1/2) orbitals. For Ca0.5, the AFM Co $e_g$ -- Ir $t_{2g}$ coupling seems to dominate due to its highly distorted structure. For Ba0.5, the increased structural symmetry weakens such hybridization but strengthen the Co $t_{2g}$ -- Ir $t_{2g}$ pathway, with the competing interactions leading to a magnetic frustration on this sample. Together with the ASD, the competing magnetic couplings explain the emergence of a SG phase at low temperatures in both compounds.

\begin{acknowledgements}
This work was supported by the Brazilian funding agencies: Funda\c{c}\~{a}o Carlos Chagas Filho de Amparo \`{a} Pesquisa do Estado do Rio de Janeiro (FAPERJ) [Nos. E-26/202.798/2019 and E-26/211.291/2021], Funda\c{c}\~{a}o de Amparo \`{a}  Pesquisa do Estado de Goi\'{a}s (FAPEG) and Conselho Nacional de Desenvlovimento Cient\'{\i}fico e Tecnol\'{o}gico (CNPq) [No. 400633/2016-7]. We thank Diamond Light Source for time on beamline I06 under proposal MM29620-1. We acknowledge DESY (Hamburg, Germany), a member of the Helmholtz Association HGF, for the provision of experimental facilities. Parts of this research were carried out at beamline P09 under proposal I-20200348 and we would like to thank J. Bergtholdt and O. Leupold for assistance in setting up the experiment and 6T/2T/2T vector magnet, respectively. 
\end{acknowledgements}

\end{document}